\documentclass[sigconf]{acmart}

\acmConference[Koli Calling '25]{25th Koli Calling International Conference on Computing Education Research}{November 11--16, 2025}{Koli, Finland}
\acmBooktitle{25th Koli Calling International Conference on Computing Education Research (Koli Calling '25), November 11--16, 2025, Koli, Finland}
\acmPrice{}
\acmDOI{10.1145/3769994.3770027}
\acmISBN{979-8-4007-1599-0/25/11}
\setcopyright{none}

\usepackage{algorithmic}
\usepackage{graphicx}
\usepackage{textcomp}
\def\BibTeX{{\rm B\kern-.05em{\sc i\kern-.025em b}\kern-.08em
T\kern-.1667em\lower.7ex\hbox{E}\kern-.125emX}}
\usepackage{lipsum}                     
\usepackage{booktabs} 
\usepackage{graphicx} 
\usepackage{tabularx}
\usepackage[english]{babel}
\usepackage{microtype}
\usepackage{hyperref}
\usepackage{soul}
\usepackage{array}       
\usepackage{pifont}      

\usepackage{tikz}
\usetikzlibrary{shapes, arrows.meta, positioning}
\usepackage{multirow}
\newcolumntype{Y}{>{\raggedright\arraybackslash}X}

\usepackage{xspace}

\newcommand{\activity}[1]{\textsc{Activity~#1}\xspace}

\begin{document}

\title{Observing Without Doing: Pseudo-Apprenticeship Patterns in Student LLM Use}

\author{Jade Hak}
\email{sjhacket@usc.edu}
\affiliation{%
    \institution{University of Southern California}
    \department{Department of Computer Science}
    \city{Los Angeles}
    \state{CA}
    \country{USA}
}

\author{Nathaniel Lam Johnson}
\email{nlj@usc.edu}
\affiliation{%
    \institution{University of Southern California}
    \department{Department of Computer Science}
    \city{Los Angeles}
    \state{CA}
    \country{USA}
}

\author{Matin Amoozadeh}
\email{mamoozad@cougarnet.uh.edu}
\affiliation{%
    \institution{University of Houston}
    \department{Department of Computer Science}
    \city{Houston}
    \state{TX}
    \country{USA}
}

\author{Amin Alipour}
\email{maalipou@central.uh.edu}
\affiliation{%
    \institution{University of Houston}
    \department{Department of Computer Science}
    \city{Houston}
    \state{TX}
    \country{USA}
}

\author{Souti Chattopadhyay}
\email{souti@usc.edu}
\affiliation{%
    \institution{University of Southern California}
    \department{Department of Computer Science}
    \city{Los Angeles}
    \state{CA}
    \country{USA}
}

\renewcommand{\shortauthors}{Hak et al.}

\begin{abstract}
    Large Language Models (LLMs) such as ChatGPT have quickly become part of student programmers' toolkits, whether allowed by instructors or not. This paper examines how introductory programming (CS1) students integrate LLMs into their problem-solving processes. We conducted a mixed-methods study with 14 undergraduates completing three programming tasks while thinking aloud and permitted to access any resources they choose. The tasks varied in open-endedness and familiarity to the participants and were followed by surveys and interviews. We find that students frequently adopt a pattern we call \emph{pseudo-apprenticeship}, where students engage attentively with expert-level solutions provided by LLMs but fail to participate in the stages of cognitive apprenticeship that promote independent problem-solving. This pattern was augmented by disconnects between students' intentions, actions, and self-perceived behavior when using LLMs. We offer design and instructional interventions for promoting learning and addressing the patterns of dependent AI use observed.
\end{abstract}

\ccsdesc[500]{Social and professional topics~CS1}
\ccsdesc[500]{Social and professional topics~Student assessment}
\ccsdesc[300]{Human-centered computing~Empirical studies in HCI}
\ccsdesc[500]{Human-centered computing~User studies}

\keywords{Large language models, computing education, help-seeking behavior, CS1, cognitive apprenticeship}

\maketitle
\footnotetext{ This work is licensed under the \href{https://firstdonoharm.dev/version/3/0/bds-bod-cl-eco-extr-ffd-law-mil-sv.html}{Hippocratic License HL3-BDS-BOD-CL-ECO-EXTR-FFD-LAW-MIL-SV}. \href{https://firstdonoharm.dev/version/3/0/bds-bod-cl-eco-extr-ffd-law-mil-sv.html}{\includegraphics[height=1em]{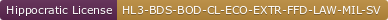}}}

\section{Introduction}
Large Language Models (LLMs) like ChatGPT and GitHub Copilot have rapidly transformed the programming landscape, including introductory computer science (CS1) courses. These tools can generate code, explain programming concepts, and debug errors on demand. Whether instructors explicitly allow LLMs or not  \cite{lau2023ban} and despite the disastrous environmental impacts of this technology \cite{morrison2025holistically}, students often integrate LLMs into their coursework.

While prior work documents patterns of AI use in computing education~\cite{kazemitabaar2023novices,amoozadeh2024student,pratherWideningGapBenefits2024}, there is limited understanding of how novice programmers engage with LLMs in real time and how these interactions shape their problem-solving strategies. In particular, little is known about how students divide cognitive work between themselves and the LLM, how much ownership they maintain over solutions, and how their attitudes toward AI influence these behaviors.

We address this gap through a mixed-methods study of 14 CS1 students performing three programming tasks of varying open-endedness and familiarity while thinking aloud. We captured detailed interaction data, including prompts to LLMs, think aloud transcripts, screen recordings, and follow-up interviews probing participants' attitudes and decision-making. Our analysis focuses on four research questions: 
\begin{itemize}
    \item \textbf{RQ1:} To what extent do students use LLMs on programming tasks?
    \item \textbf{RQ2:} How do students prompt LLMs during programming tasks?
    \item \textbf{RQ3:} How do students work with LLM outputs after prompting?
    \item \textbf{RQ4:} What are the attitudes and perceptions of students toward using LLMs on programming assignments?
\end{itemize}

Our findings reveal a pattern we call \emph{pseudo}-\-\emph{apprenticeship}: students treating LLMs as expert models, bypassing planning and exploration to work backward from complete solutions. This pattern resembles the early modeling phase of cognitive apprenticeship~\cite{collins1989cognitive}, but without the guided transition toward autonomy.

This paper makes the following contributions:
\begin{enumerate}
    \item A characterization of how novice programmers integrate LLMs into their problem-solving workflow across tasks with varying open-endedness and familiarity.
    \item Evidence of a pseudo-apprenticeship learning pattern, where students rely heavily on LLM-generated solutions but engage minimally in refinement or independent reasoning.
    \item Implications for pedagogy, including strategies for designing assignments and instructional interventions that maintain learning benefits while mitigating overreliance on AI.
\end{enumerate}
These findings contribute to ongoing conversations about how computing education can adapt to support learners in light of the advent of generative AI.

\section{Background and Related Work}
\subsection{Generative AI in Programming Education}

The use of Large Language Models (LLMs) in programming education has rapidly evolved, trending toward widespread adoption. Early work examined institutional policies and instructor responses~\cite{lau2023ban,sheard2024instructor}. More recent research has examined structured approaches to introducing AI into courses, such as debugging assistants and tutor-like interfaces~\cite{liu2024teaching,sheese2024patterns}. A systematic review by Raihan et al.~\cite{raihan2025large} synthesized 125 studies of LLMs in computer science education, noting that most studies emphasized tool capabilities and performance benchmarks, with less focus on how students actually use LLMs during learning.

\subsection{Patterns of LLM Use Among Students}

Research examining student interactions with LLM-based programming support has found that many students focus on immediate, task-specific assistance rather than conceptual understanding. Sheese et al.~\cite{sheese2024patterns} observed that most queries targeted debugging errors or implementing assignment solutions, often with minimal contextual information. Similarly, Piscitelli et al.~\cite{piscitelli2024influence} reported that chatbot access reduced task completion time and errors but that many students requested full solutions instead of incremental checks or debugging guidance. Penney et al.~\cite{penney2025outcomes} found that ChatGPT users tended to issue brief, zero-shot prompts with little refinement compared to peers working with human tutors.

Amoozadeh et al. \cite{amoozadeh2024student} conducted a study of 15 CS1 students using ChatGPT, finding that one-third of participants prompted for complete solutions before making an effort on their own, and that students often failed to verify the correctness of AI-generated code. Shah et al.~\cite{shah2025students} observed upper-division students using GitHub Copilot for large codebase tasks, identifying a pattern of 'one-shot prompting' where students requested complete feature implementations and then cycled through debugging or regeneration requests.

\subsection{Student Perceptions}

Computer science students reflect mixed perceptions of LLMs and their place in learning. Penney et al.~\cite{penney2025outcomes} reported that while students distrusted ChatGPT's accuracy for learning new concepts, they felt more comfortable asking it basic questions than approaching human tutors, largely because it reduced fear of judgment. Lyu et al.~\cite{lyu2024evaluating} similarly observed that students valued the LLMs' abilities to understand questions and assist with syntax but raised concerns about their limited support for critical thinking and increasingly favored human teaching assistants over time.

\subsection{Theoretical Perspectives on Student–LLM Interactions}

Established learning theories help explain how students might engage with LLMs. Vygotsky's Zone of Proximal Development (ZPD)~\cite{vygotsky1978mind} highlights how temporary support can enable learners to accomplish tasks they cannot yet complete independently. Scaffolding~\cite{wood1976role} structures this support so that it can be gradually withdrawn as learners gain competence, a principle embedded in many intelligent tutoring systems~\cite{zhang2021ai}.

Help-seeking research offers a complementary lens. Nelson-Le Gall~\cite{nelson1981help} distinguished between instrumental help-seeking, where learners request hints or partial information that supports independent problem-solving, and executive (or expedient) help-seeking, where learners seek complete answers to minimize effort. Adaptive help-seeking adds a metacognitive dimension, emphasizing when to seek help, what type of help to request, and how to use it productively~\cite{aleven2003help,aleven2016help,roll2011improving}. While these frameworks are well established in traditional educational contexts, their application to open-ended, on-demand LLM use is still emerging.

\subsection{Positioning Our Work}

Most prior work has observed students interacting with LLMs in constrained settings, such as single task types, mandated AI tools, or structured tutoring interfaces~\cite{amoozadeh2024student,piscitelli2024influence,penney2025outcomes}. Our study differs in three ways: (1) we vary task familiarity and openness, ranging from well-known algorithmic exercises to open-ended creative design; (2) we allow students to freely choose any resources, including LLMs, web search, notes, or none at all; and (3) we create a naturalistic environment similar to homework, with no academic integrity enforcement. This design captures how students integrate LLMs into their workflow under realistic, unconstrained conditions.

\section{Methodology}\label{sec:methodology}
This section covers how we recruited participants, biographical information on the participants, and our study protocol.

\subsection{Recruitment}
We recruited participants by advertising the study in two introductory computer science courses at the University of Southern California (USC): CSCI 102, Foundations of Computer Science (equivalent to CS0) and CSCI 103, Introduction to Programming (equivalent to CS1). Recruitment took place both in class and through posts on the corresponding Piazza discussion boards. Participants were informed that the study had no impact on their grades and were compensated \$30 for their time. The study was approved by the University's IRB as of minimal risk to participants.

We selected participants who used LLMs on recent programming assignments to observe patterns among regular users. We also screened for participants who were early in learning computer science at an undergraduate level. Among 390 students, 57 responded (14.6\% response rate). Of the 57 respondents, we excluded one respondent under the age of 18, three respondents who had taken more than two years of computer science courses, and four respondents who had never used AI tools, like ChatGPT. We excluded a further 29 respondents who reported not having used AI tools, like ChatGPT, on at least some of their computer science assignments in the last three months. Of the 20 eligible respondents, 14 respondents scheduled a time to participate and completed the study.

\subsection{Study Protocol}
Our study protocol consists of a pre-survey, a set of three programming activities, a post-survey, and an interview. Each participant completed the study individually in a single one-hour session, under the direction of a study investigator. The study was completed via Zoom video calls, which were recorded with the consent of the participant and investigator. Participants shared their screens during the programming activities and thought aloud through said activities.

\subsubsection{Pre-Survey}

Each participant completed a pre-survey at the beginning of the study. The pre-survey consisted of the following sections that we asked them questions about: demographics, programming experience, academic history, self-efficacy, problem-solving approaches, preferred methods of learning, use of AI tools, and attitude toward AI tools.

\subsubsection{Programming Activities}
Participants were asked to complete three programming activities in the C++ programming language using Codio\footnote{\url{https://www.codio.com/}}, a web-based programming environment.
We instructed the participants to treat the activities as their CS course assignments and that they could use any web resources, including LLMs, class notes, and web search, to the extent they desired. We also informed the participants that they had five, 10, and 30 minutes for activities \activity{1}, \activity{2}, and \activity{3}, respectively. However, in case the participants were making significant progress at the expiration of the time limit, we allowed additional time for participants to wrap up and prompted them to move to the next task. Participants were asked to think aloud while completing the tasks and were given a sample task to think aloud while working on before starting the programming activities.

\activity{1}, \activity{2}, and \activity{3} were ordered in increasing conceptual complexity and scope. They were presented in the same order to all participants due to this ordinal characteristic.

\paragraph{\activity{1}: Find the Maximum Number}
Participants were asked to \emph{write a program that finds the maximum number in a series of inputs. The program should prompt the user to enter a series of numbers and then output the maximum number entered.} We expected that all CS0 and CS1 students would have the required knowledge to complete this task. Through this task, we observed how participants work in the presence of LLMs when faced with a familiar problem.

\paragraph{\activity{2}: Countdown Timer}
Participants were asked to \emph{write a program that prints integer numbers from 10 to zero, decreasing by one each second. The program should use a delay to ensure that each number is printed one second apart.} We expected that participants would likely be unfamiliar with how to implement a `delay' function, as this was beyond the concepts covered in the CS0 or CS1 curriculum. Through this task, our objective was to observe how participants work in the presence of LLM when faced with unfamiliar tasks that require learning new concepts.

\paragraph{\activity{3}: Game}
Participants were asked to \emph{design and implement a game.} We selected a loosely defined task to allow participants to engage in both design and implementation. The open-endedness of the task, in terms of requirements, scope, and implementation, allowed participants to be as creative as they wanted. Participants were instructed that their programs for this activity would be evaluated based on functionality, usability, creativity, and code quality. The objective of this task was to observe how participants work in the presence of LLMs when facing a broadly defined development task that required them to ideate, plan, and implement under a limited time budget.

\subsubsection{Post-Study Survey}

After the programming activities, we used a survey to gather participants' reflections on their use of LLM during the activities. Participants who did not use an LLM during programming activities were excluded from this survey. The survey aimed to investigate how the participants' use of AI during programming activities affected, among other factors, their motivation to complete the activities, their learning during the activities, and the quality of their code. The survey also asked participants to evaluate the reliability and usefulness of the LLM's responses.

\subsubsection{Interview}

At the end of each study session, we conducted a ten-minute semi-structured interview to gain insights into participants' experiences during the study session and perspectives on LLMs. The interview focused on three key areas: (1) understanding of the code they had written, (2) how they approached and resolved challenges during the session, and (3) their attitudes toward large language models in the context of programming and learning. Example questions included: ``Were there any moments when you felt stuck while coding? If so, how did you overcome them?'', ``How did you decide when to use an LLM?'', and ``How do you feel about using LLMs for coding tasks?''

\subsection{Participants}
14 participants completed our study. Ten participants were first-year students, three were second-year students, and one was a third-year student. All but two of the participants were majoring in a form of computer science. Nine participants identified as female, four as male, and one participant preferred not to disclose their gender.

Our participants reported a wide range of prior programming experience; one participant reported only one month of programming, while two participants had four years of experience. Our participants are all undertaking early and intermediate CS courses, having either completed or are currently enrolled in undergraduate CS courses.

The participants reported varying frequencies of LLM use.
When asked about their typical frequency of LLM usage over the past three months, 1 participant reported using LLMs a few times a month, 8 reported using LLMs a few times a week, and 5 reported daily usage. All participants reported using LLMs for computer science assignments in the past three months. 11 and 3 participants reported using LLMs for some and most assignments, respectively.

Half of the participants reported debugging as their reason for using LLMs on programming assignments. When participants were questioned about strategies the participants use to verify AI-generated code, the most common response was to "run and test the code", followed by "review code line by line" and "compare [code] with documentation".

\section{Data Analysis Methods}\label{sec:data-analysis-methods}
\subsection{Programming Activities}

\subsubsection{Segmentation and Annotation}
Our data from the programming activities consists of 14 video sessions, one from each participant, with accompanying transcripts. We analyzed this data by segmenting the data into short time segments, labeling the participants' behaviors during each such time segments, and inductively developing and applying codes to categorize the participants' behavior during the time segments. Two investigators collaboratively worked on this segmenting, annotating, and coding process.

To segment the videos, we start and end each segment when the participant's actions indicate a shift in focus or goal. We attained 552 segments, averaging 49 seconds in length each, with an average 39 segments per participant.

For each time segment, the investigators labeled the following:

\begin{itemize}
    \item Actions: What the participant does
    \item Goal: What the participant is working toward
    \item Verbalizations: Anything the participant says
    \item AI Tools Used: Any AI tools used, prompts for such tools
    \item Non-AI Resources Used: Any lecture notes, web searches, or other resources used
\end{itemize}

\subsubsection{Inductive Coding}

To understand how participants made use of LLMs, we inductively coded: (1) when participants were interacting with an LLM, (2) how participants were interacting with LLM, and (3) the participants' intentions for prompting with an LLM. Two authors conducted this inductive coding collaboratively. We used the segment-level annotations, considering a participant's actions, verbalizations, and LLM prompts, at the time of and leading up to an LLM interaction.

Our inductive coding produced the following four categories of \textit{interacting with an LLM}: (1) prompting an LLM, (2) reading a response from an LLM, (3) copying code from an LLM into one's IDE, or (4) integrating LLM-generated code into one's own code. We labeled occurrences of these interactions, allowing us to determine when participants were interacting with an LLM.

In further breaking down how participants used responses from LLMs, we obtained three broad categories of use of LLM responses: (1) coding, (2) comprehending, and (3) ideating. Within each category we also inductively coded subcategories of use, which can be found in our supplementary materials. We also inductively coded the manner in which participants used responses from an LLM. We developed the following categories of \textit{manner of using an LLM response}: (1) coding, (2) comprehending, (3) ideating, and (4) other.

Finally, we wanted to code why participants prompted LLMs. This intention was informed primarily by the prompt itself, alongside the participant's verbalizations and history of actions leading up to the prompting. We initially developed 11 intentions for LLM use. We distilled these labels to six broader categories of intention for LLM use. Table~\ref{table:llm_intentions} presents the final categories, associated subcategories, definitions, and representative prompts.

\begin{table*}[htbp]
    \caption{Inferred Intention for Prompting LLM}
    \label{table:llm_intentions}
    \centering
    \begin{tabularx}{\linewidth}{@{}lX X X@{}}
        \toprule
        \textbf{Category}                  & \textbf{Definition}                                                             & \textbf{Example Prompt}                                                                                     \\
        \midrule
        Algorithm and Planning             & Asking an LLM to design an algorithm and/or plan an approach to the problem     & Ok I'm going ahead with the Math Quest option. Do you have suggestions on how to get started coding in C++? \\
        Code Explanation                   & Asking an LLM to explain how a piece of code works                              & Can you explain it more simpler [sic], I don't understand the 5th step                                      \\
        Brainstorming                      & Asking the LLM to come up with ideas for approaching a problem                  & Give me simple game ideas                                                                                   \\
        Conceptual and Syntactic Questions & Asking the LLM to answer conceptual or syntactic questions                      & What does std:: do?                                                                                         \\
        Debugging               & Asking the LLM to identify or fix errors (e.g. compilation, runtime, test cases) in code           & \texttt{\{[}CODE{] \}} why my code failed when the output should be 2.14748e+09                                                                                                               \\
        Code Writing                       & Using the LLM to generate chunks of code                                        & Use a timer to implement this and keep it simple                                                            \\
        \bottomrule
    \end{tabularx}
\end{table*}

\subsection{Interview Data}

We conducted qualitative analysis of our interview data. This analysis process was conducted collaboratively by two of the authors. Working with the interview transcripts, we first unitized the data by extracting meaningful segments that captured distinct ideas or responses. This unitization process yielded 407 snippets across all participants, with an average length of 17 words per segment.

Following unitization, we conducted open coding on these snippets. Each snippet was assigned a code that captured its meaning or content. This initial coding generated 139 unique codes, representing the variety of ideas and reflections presented in the interviews.

We then proceeded with a multi-stage thematic analysis to identify patterns across the coded data. This involved an iterative process of successively grouping related codes.

During this analytical process, we excluded codes that were tangential to our research questions or that did not demonstrate sufficient patterns across multiple participants.

From the 139 unique codes, we successively organized the codes into 100, then 20, then 10, and finally 4 broader themes. Our themes and further analyses are presented in Section~\ref{sec:results}.

\section{Results}\label{sec:results}
Except for P13, each participant worked on all three activities until completion or until being interrupted by the researcher for lack of time. P13 stopped their participation 5 minutes into each \activity{2} and \activity{3}, overwhelmed by the difficulty of the coding tasks, but continued to the post-survey and interview. Using our data analysis, we answer the research questions:

\subsection{\textbf{RQ1}: To what extent do students use LLMs on programming tasks?}\label{subsec:Results-RQ1}
    
While 64\% (9/14) of participants used LLMs for \activity{1}, 86\% (12/14) used them for each \activity{2} and \activity{3}, aligning with our hypothesis that familiar concepts reduce LLM reliance.

The extent of AI use varied across participants and activities.


Participants prompted LLMs an average of 5.9 times per session (median=5.5), ranging from zero (P5) to 15 prompts (P11). P11's extensive prompting reflected a strategy of iteratively refining AI-generated code. P5 did not engage with LLMs while P3 only engaged with LLMs through AI generated results in web search.

\begin{figure}[htbp]
	\centerline{\includegraphics[width=\columnwidth]{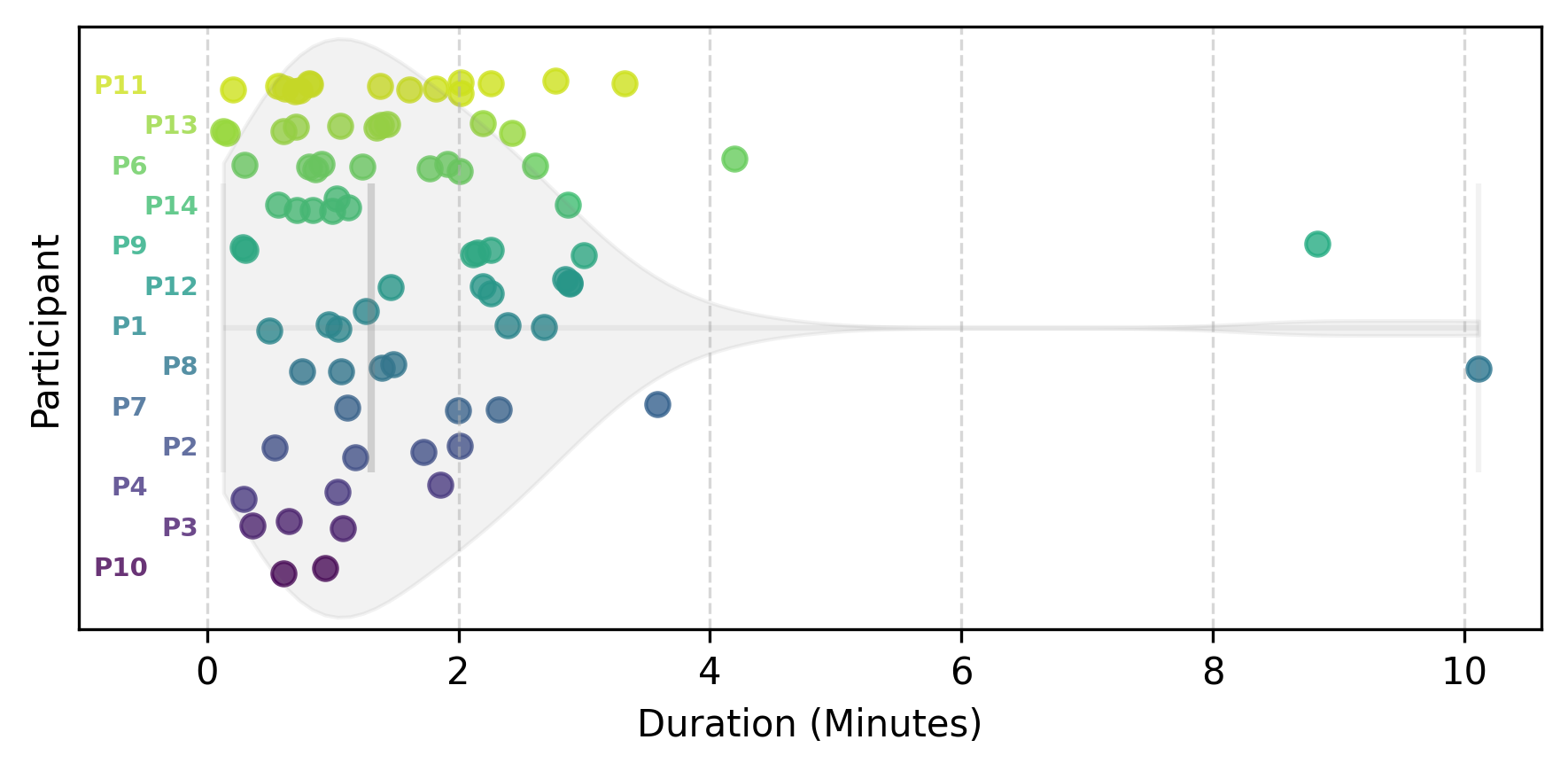}}
	\caption{Time Spent Per Prompt, by Participant. Participants are ordered by number of prompts.}
	\label{fig:time_per_prompt_by_participant}
\end{figure}

Using the definition and labeling of \textit{interacting with an LLM} from Section~\ref{sec:data-analysis-methods}, we analyzed the time participants spent interacting with an LLM for each prompt, seen in Figure~\ref{fig:time_per_prompt_by_participant}.

Some participants spent extended periods with single prompts. P8 spent 10 minutes iteratively implementing LLM-suggested improvements, while P9 spent 9 minutes working with an unsolicited complete solution the LLM provided in response to a request for ``getting started'' guidance.

\subsubsection{Duration of LLM Interaction}

\begin{figure}[htbp]
	\centerline{\includegraphics[width=\columnwidth]{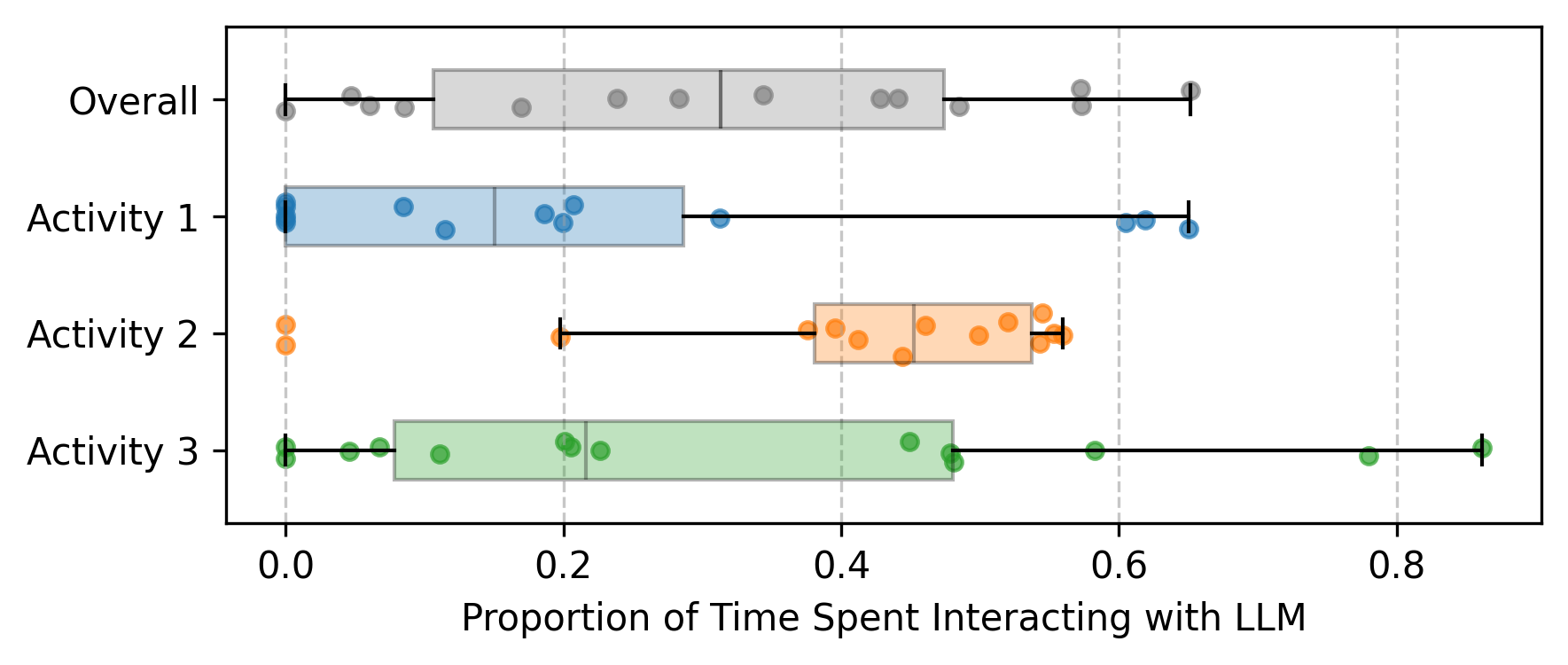}}
	\caption{Time Spent Interacting with LLM, by Activity.}
	\label{fig:box_proportion_time_spent_all}
\end{figure}

Figure~\ref{fig:box_proportion_time_spent_all} shows the distribution of time spent interacting with an LLM, by activity. Across all activities, the students dedicated approximately one-quarter of their time to LLM interactions, demonstrating that these tools comprised a substantial component of their programming workflow.

In \activity{1} participants displayed more independence than the other Activities, with LLM interactions occupying only about 10\% of their time, though some students still sought substantial AI guidance. In contrast, when faced with unfamiliar programming concepts in \activity{2}, the students leaned heavily on LLM assistance, spending nearly 40\% of their time engaged with these tools.

During the open-ended \activity{3}, we observed the widest variation in LLM engagement. Our screen recordings revealed distinct approaches: 4 participants used LLMs as an initial brainstorming tool, 5 participants used LLMs for short, syntax-based questions alone, 2 participants relied heavily on the AI for code writing, and, unlike the previous activities, no participant used LLMs for debugging alone.

\subsubsection{Debugging Strategies}

Despite 43\% self-reporting IDE debugger use in the pre-survey, no participants used debuggers during tasks. Instead, they relied on LLMs (57\%) and print statements (36\%), revealing a gap between reported and actual debugging practices.

All participants except for two (12/14) reported turning to an LLM at least sometimes when encountering a problem with their code, with 4 (29\%) of these participants turning to LLMs most times in these instances. Further, half of participants identified debugging as their primary motivation for using LLMs at all on programming tasks.

These patterns indicate that LLMs play a key role in the participants' debugging workflows. The discrepancy between the pre-survey responses and the observed behavior during the programming activities suggests that the participants may rely on LLMs for debugging more than they realize.

\subsection{\textbf{RQ2}: How do students prompt LLMs during programming tasks?}\label{subsec:Results-RQ2}

To analyze how participants divided cognitive work between themselves and LLMs, we look at (1) when participants first engage an LLM, (2) the amount of context participants provide to the LLM about the problem, and (3) the types of responses requested from the LLM.

\subsubsection{Timing of First Prompt}\label{subsec:RQ2-Extent-Individual-Work}

\begin{figure}[htbp]
	\centerline{\includegraphics[width=\columnwidth]{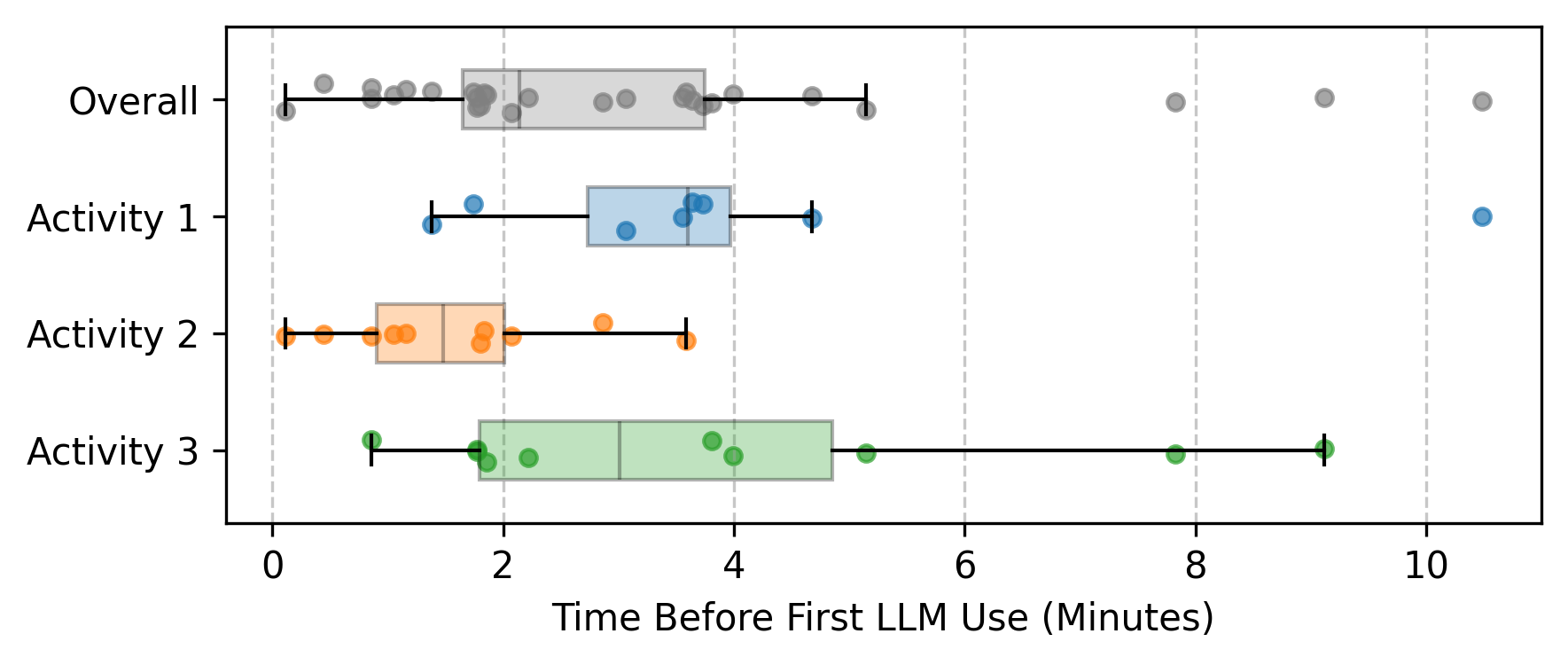}}
	\caption{Time Spent Before First LLM Use (excludes P7, \activity{2} where participant searched web before later using LLM)}
	\label{fig:box_plot_first_llm_use}
\end{figure}

Figure~\ref{fig:box_plot_first_llm_use} shows time-to-first-prompt for each activity. Participants delayed prompting in \activity{1}, typically writing and compiling code before seeking help. In \activity{2}, unfamiliar threading concepts prompted earlier use, with many students consulting LLMs before coding. \activity{3}produced the widest range of timing strategies, from early brainstorming to later debugging support, reflecting its open-ended design.

\subsubsection{Context Shared with LLMs}

We examined the problem context participants included in their LLM prompts, focusing on how timing and scope shaped the LLM's role. Minimal context, such as syntax-only questions, limited responses to narrow technical help, while full context positioned the LLM as a co-designer or lead problem-solver.

Individual prompts reflect the effect of context on the LLM's role in problem solving. P6's isolated query, “are you able to output colored text in terminal,” withheld the game context and constrained the LLM to a narrow answer. By contrast, P1 shared the full specification while asking, “do not give me the answer but teach me how to do it,” yet still received complete code and shifted their work from problem-solving to solution comprehension.

\begin{figure}[htbp]
	\centering
	\includegraphics[width=0.8\columnwidth]{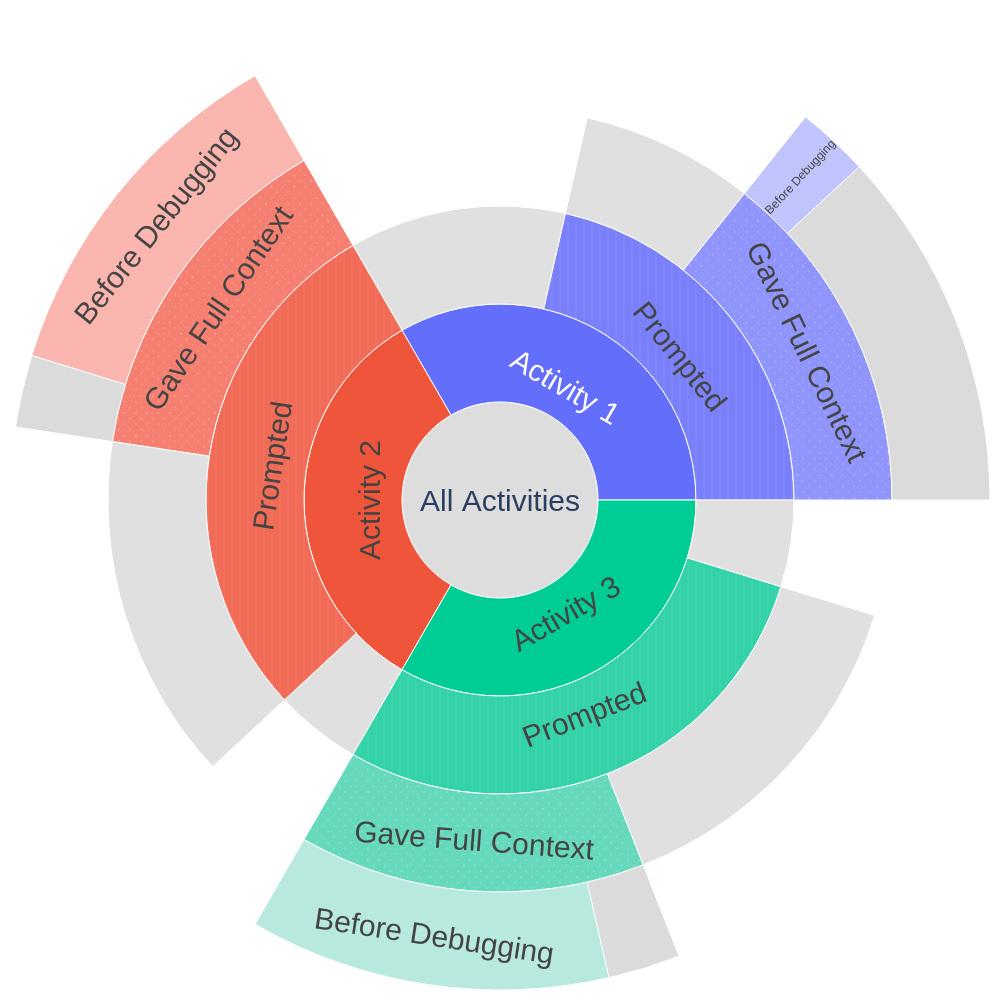}
	\caption{Information Inclusion in Prompts}
	\label{fig:information_inclusion_in_prompts}
\end{figure}

Figure~\ref{fig:information_inclusion_in_prompts} summarizes the proportion of participants who (1) prompted an LLM at all, (2) included problem context in any prompt, and (3) shared context before debugging. Early context sharing occurred more frequently in \activity{2} and \activity{3} than in \activity{1}. In \activity{1}, most participants attempted a solution independently before prompting, and only one provided problem context early. In \activity{2}, the threading-based countdown timer led several participants unfamiliar with threading to paste or paraphrase the specification before coding. In \activity{3}, the open-ended game design task encouraged some to begin by asking for game ideas, effectively giving full context even when they did not paste the specification.

\subsubsection{Prompting Intentions}

\begin{figure}[htbp]
	\centerline{\includegraphics[width=\columnwidth]{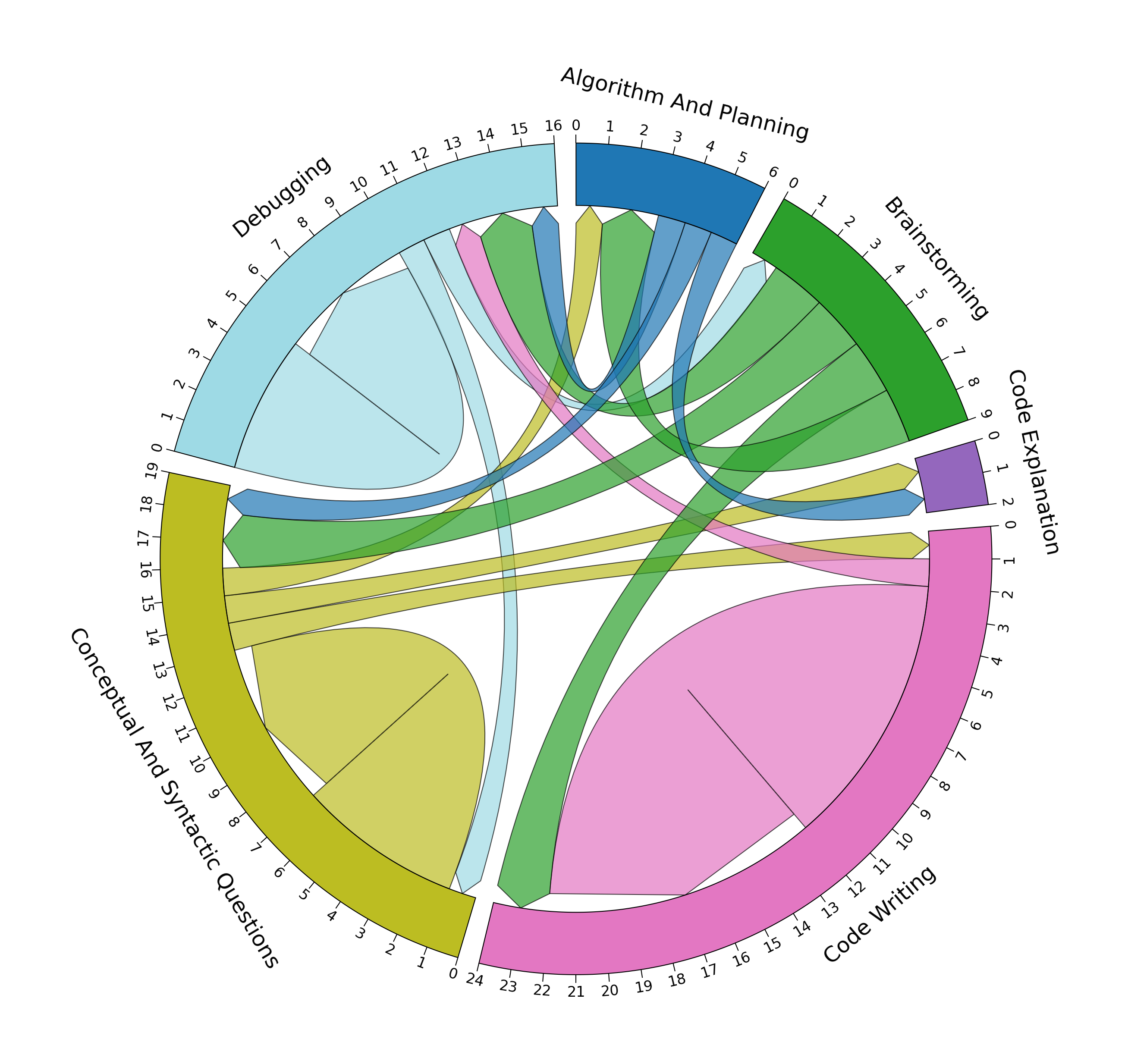}}
	\caption{Flow of Prompting Intentions: From Initial to Subsequent LLM Use. Arrows represent transitions from a participant's initial prompting intention to all subsequent intentions within the same activity. Tick marks indicate individual participant-activity pairs. Loops show cases where only one intention was used throughout an activity.}
	\label{fig:first_llm_use_later_llm_use}
\end{figure}

We coded prompts based on what participants sought from the LLM, grouping them into six intentions: code writing, conceptual or syntactic questions, debugging, brainstorming, algorithm/planning, and code explanation (Table~\ref{tab:llm_prompt_participants_total}). Brainstorming was a common entry point (21\% of participant-activity pairs) but was never used alone, always transitioning into another prompt intention (Figure~\ref{fig:first_llm_use_later_llm_use}). Half of all sessions showed a single primary intention, most often code writing. Code writing saw little interplay with other prompt intentions. Just one participant moved away from code writing prompts after starting there and in only 7\% (3 of 42) of participant-activity pairs shifted from an initial non-code-writing intention to code writing.

\begin{figure*}[htbp]
	\centering
	\includegraphics[width=\textwidth]{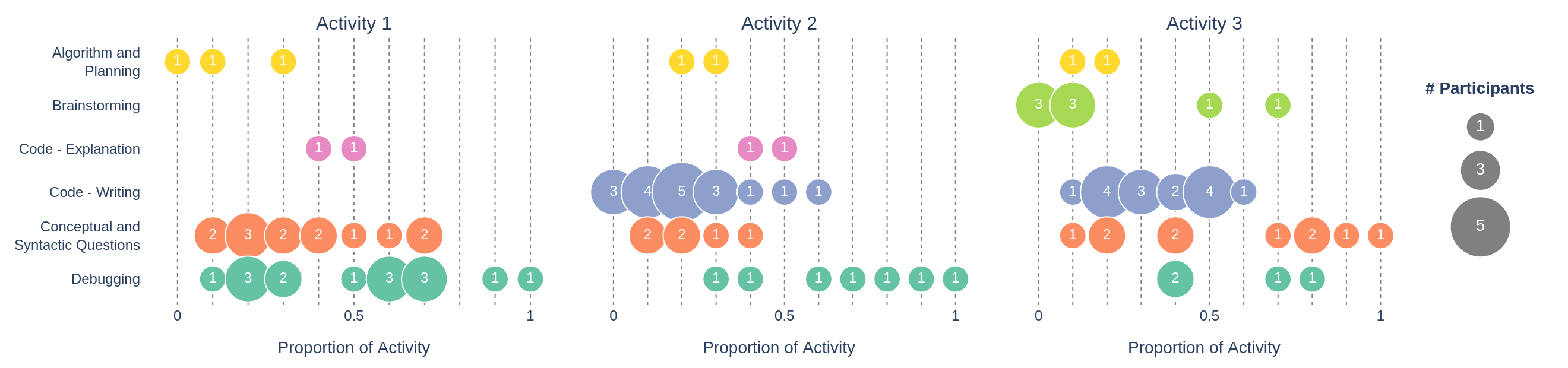}
	\caption{Intention of LLM Prompts By Activity}
	\label{fig:llm_intention_network_all}
\end{figure*}

Prompting intentions also varied by activity (Figure~\ref{fig:llm_intention_network_all}). In \activity{1}, no participants used an LLM for code writing, likely reflecting its similarity to tasks already practiced in coursework. In contrast, \activity{2} saw early and frequent code writing prompts, consistent with its unfamiliar threading requirements, while \activity{3} often began with brainstorming about game ideas before transitioning into algorithm or planning support.

\begin{table}[htbp]
	\centering
	\caption{Frequency of LLM Prompting Intentions}
	\label{tab:llm_prompt_participants_total}
	\newcolumntype{L}[1]{>{\raggedright\hangindent=1em\hangafter=1\arraybackslash}p{#1}}
	\newcolumntype{C}[1]{>{\centering\arraybackslash}p{#1}}
	\begin{tabularx}{\linewidth}{@{} L{3.5cm} C{1.8cm} C{2.3cm} @{}}
		\toprule
		\textbf{Intention Category}        & \multicolumn{2}{c}{\textbf{Frequency}}                       \\
		\cmidrule(l){2-3}
		                                   & \textbf{\# Participants}               & \textbf{\# Prompts} \\
		\midrule
		Code Writing                       & 9                                      & 27                  \\\addlinespace
		Conceptual and Syntactic Questions & 8                                      & 23                  \\\addlinespace
		Debugging                          & 8                                      & 17                  \\\addlinespace
		Brainstorming                      & 5                                      & 8                   \\\addlinespace
		Algorithm and Planning             & 4                                      & 5                   \\\addlinespace
		Code Explanation                   & 2                                      & 3                   \\
		\bottomrule
	\end{tabularx}
\end{table}

\subsubsection{Requests for Full Solutions}

In addition to the six prompt intention categories of Table~\ref{table:llm_intentions}, we separately identified instances where participants' prompts sought complete solutions to the task at hand. Seventy-nine percent of participants requested a full solution at least once, often before writing code in \activity{2} and \activity{3}.

These requests were not always explicit, such as prompting ``solve this problem'' but were evident from context, such as pasting the full problem specification without further instruction. This is an action that, given typical LLM behavior, is functionally equivalent to requesting a complete solution.

\subsection{\textbf{RQ3}: How do students work with LLM outputs after prompting?}\label{subsec:Results-RQ3}

\begin{figure*}[htbp]
	\centering
	\includegraphics[width=\textwidth]{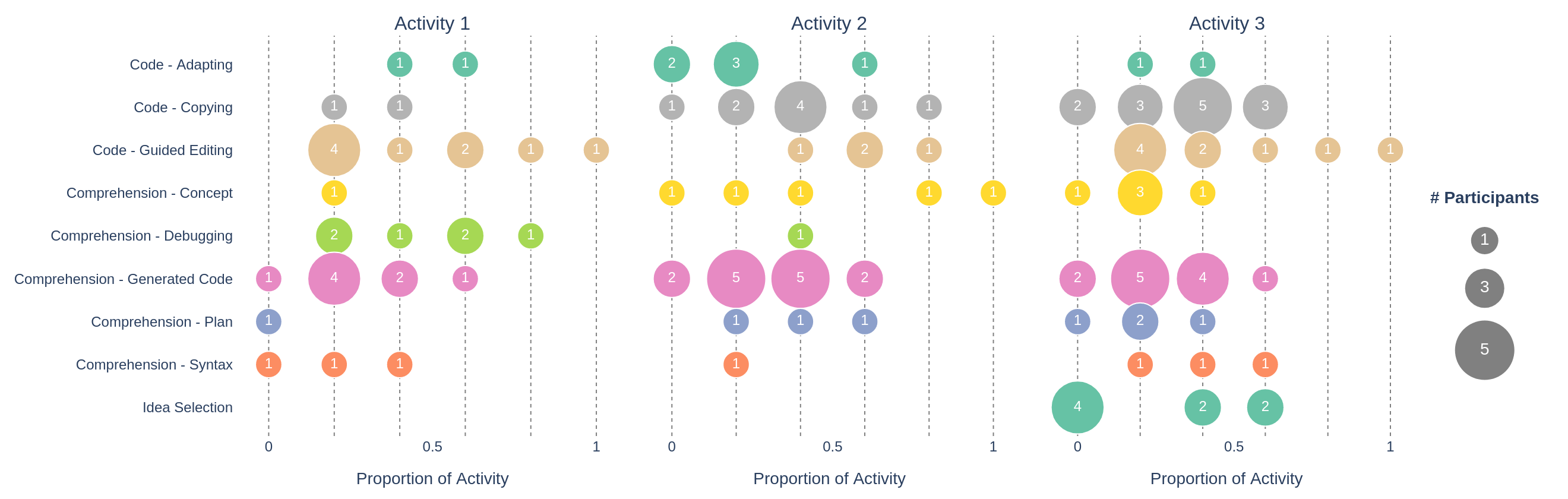}
	\caption{Actions With LLM Responses By Activity}
	\label{fig:llm_use_network_all}
\end{figure*}

Participants engaged with LLM outputs across three broad categories of action (Table~\ref{tab:llm_action_duration_participants}):
(1) \textbf{Code work}, including copying output directly, adapting it, or performing guided editing;
(2) \textbf{Comprehension}, such as reading generated code, interpreting debugging explanations, or reviewing concepts, plans, or syntax;
and (3) \textbf{Idea selection}, where students evaluated approaches or design options. Figure~\ref{fig:llm_use_network_all} shows how these actions unfolded over time by activity.

\begin{table}[htbp]
	\centering
	\newcolumntype{L}[1]{>{\raggedright\hangindent=1em\hangafter=1\arraybackslash}p{#1}}
	\newcolumntype{C}[1]{>{\centering\arraybackslash}p{#1}}
	\caption{Number of participants and mean time spent on each LLM action.
		Mean time (in minutes) is averaged only over participants who performed that action.}
	\label{tab:llm_action_duration_participants}
	\begin{tabularx}{\linewidth}{@{} L{3cm} C{1.8cm} C{3cm} @{}}
		\toprule
		\textbf{LLM Action}    & \textbf{\# Participants} & \textbf{Mean Duration Per Participant (minutes)} \\
		\midrule
		\textbf{Code}          &                          &                                                  \\
		\quad Adapting         & 7                        & 0.80                                             \\
		\quad Copying          & 11                       & 1.43                                             \\
		\quad Guided Editing   & 7                        & 2.78                                             \\
		\addlinespace
		\textbf{Comprehension} &                          &                                                  \\
		\quad Concept          & 4                        & 1.64                                             \\
		\quad Debugging        & 4                        & 1.13                                             \\
		\quad Generated Code   & 8                        & 3.93                                             \\
		\quad Plan             & 3                        & 1.75                                             \\
		\quad Syntax           & 5                        & 0.98                                             \\
		\addlinespace
		Idea Selection         & 5                        & 1.68                                             \\
		\bottomrule
	\end{tabularx}
\end{table}

Most participants opted for direct copying of LLM-generated code, with little time spent adapting this code. However, those who engaged in comprehension activities, also more than half of participants, spent considerably longer on these tasks than any other LLM-related actions, suggesting these students were intentional in understanding the AI-generated content.

Figure~\ref{fig:llm_use_network_all} shows how these actions unfolded over time by activity. Participants spent time understanding code generated by the LLM during the initial 60\% of each activity.

\subsubsection{Mismatch Between Prompt Intentions and Actions}

Participant actions often diverged from their original prompting intentions. Conceptual or brainstorming prompts frequently returned complete code solutions, shifting students from exploratory or planning work to direct code adoption. For example, one participant asked for “other game options” but received a fully implemented game and integrated it immediately:

\begin{quote}
	``It just gave me a full game. I wasn't expecting that, but I used most of it because it was already done.'' (P1)
\end{quote}

\begin{figure}[htbp]
	\centerline{\includegraphics[width=\columnwidth]{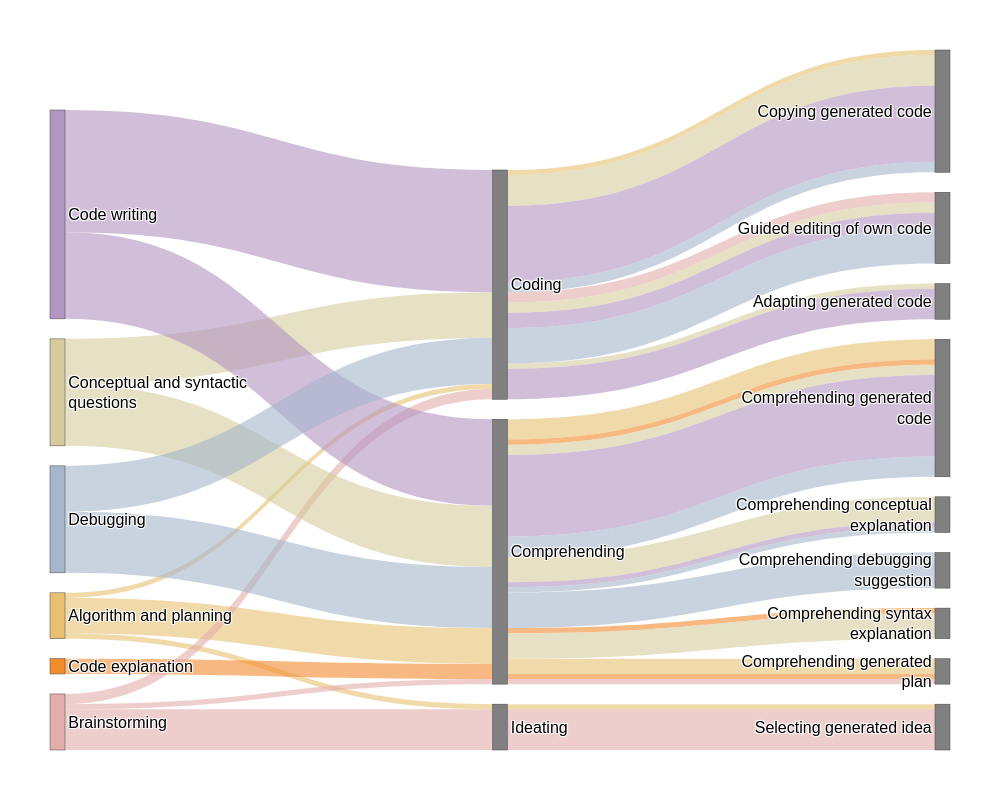}}
	\caption{Mapping of LLM prompt intentions to subsequent participant actions.
		Arrows indicate the flow from prompt intention to action category to action subcategory.
		Flow width represents the total number of prompts leading to each type of action, aggregated across all participants.}
	\label{fig:sankey_count_all_activities}
\end{figure}

\sloppy
Figure~\ref{fig:sankey_count_all_activities} maps prompting intentions to subsequent actions. Prompts seeking code generation most often led to comprehension or direct copying of generated code, showing alignment between intention and outcome. Debugging prompts typically produced guided edits of participants’ own code rather than wholesale adoption of generated solutions. In contrast, planning and algorithm prompts often triggered unsolicited, full implementations, shifting participants toward comprehending completed solutions rather than reviewing or refining plans.

\subsection{\textbf{RQ4}: What are the attitudes and perceptions of students towards using LLMs on programming assignments?}\label{subsec:Results-RQ4}

In the interviews, participants described using LLMs less as a deliberate learning strategy and more as a way to manage the practical pressures of coursework. Their attitudes often conflicted with their actual practices, revealing tensions between speed, understanding, and academic norms. The following subsections outline their motivations, perceived learning trade-offs, trust and validation practices, and the disconnects between self-reported and observed use, as gleaned from our thematic analysis of the interview data.

\subsubsection{Motivations and Usage Triggers}

Time pressure emerged as the dominant driver of LLM use. Nine participants (P1, P2, P3, P4, P7, P10, P11, P12, P14) cited efficiency as their primary motivation, even when understanding the problem without AI (P6, P8, P9, P11, P12). Participants reported preferring lecture notes, videos, and human help over LLMs when focused on learning (P2, P3, P5, P6, P10, P14) but valuing the immediacy of support from LLMs when working on assignments.

Task complexity shaped usage decisions. Some participants reported turning to LLMs for frustrating but minor errors like segmentation faults (P2, P5, P7, P8), while others avoided LLMs for tasks deemed too simple: ``I haven't used it in my C++ class because the labs are pretty easy'' (P13). Participants also used LLMs to bridge knowledge gaps, with P1 explaining: ``The way information is originally presented doesn't make sense to me, so I ask AI to simplify it.''

\subsubsection{Learning Trade-offs}

Participants described mixed effects of LLM use on learning. Some reported incidental learning, particularly picking up new syntax when reading AI-generated code (P2, P6, P7, P9, P12, P14). Others felt that relying on LLMs bypassed opportunities for critical thinking (P9, P11).

Two participants prioritized the long-term learning that they assert comes from avoiding LLM use (P10, P13). P13 explained, ``I try to work through problems on my own first because I won't always have ChatGPT available.'' In contrast, P11 expressed, ``I would absolutely prioritize my grade over my learning. If that meant using AI to complete the work quickly, I would.''

Some reported avoiding LLMs when prohibited (P2, P4), while others described using LLMs despite feelings of guilt for hindering learning. P10 articulated a collective dilemma driving LLM use, stating, ``Everyone is getting their assignments in on time, and it can feel like I'm at a disadvantage if I don't use [an LLM].''

\subsubsection{Trust, Control, and Validation}

All participants rated LLMs as at least somewhat reliable in post-surveys, and ten noted in interviews that LLMs occasionally produced errors. Several participants found that LLM responses contained irrelevant information (P3, P7, P8, P10) or were overwhelming (P2, P14). Trust varied by task type, with participants expressing confidence in LLMs for simple syntax (P2, P4, P7, P9) but skepticism about complete solutions.

Most participants (8/14) described critically evaluating AI-generated codes. P9 noted, ``I reviewed all the functions and lines to make sure I understood what was happening.'' Two described cross-checking LLM outputs against non-AI resources (P1, P6). However, two participants (P11, P14) described a copy-first approach. P14 reflected, ``I copied and pasted [the generated code] because I had no idea how to solve the problem...I do not know what thread::sleep\_for means.''

Six participants (P6, P8, P9, P10, P11, P12) emphasized maintaining intellectual ownership, using LLMs as supplements rather than replacements for their own reasoning. P12 said, ``I already knew what I wanted to do, and I used ChatGPT to help me execute that approach.'' Yet several acknowledged moments when they ceded ownership, sometimes unintentionally. P9 admitted, ``It was slightly disappointing that [the LLM] gave me everything, but at the same time, it was convenient.''

\subsubsection{Self-Reported vs. Actual Use}

Interviews revealed gaps between participants' stated workflows and their observed behaviors. Many claimed to use LLMs only after developing their own solution or exhausting non-AI methods (P2, P3, P4, P5, P6, P7, P10, P12). This included hierarchies like checking compiler errors, then turning to debuggers or peers, and finally using LLMs as a ``last resort'' (P2, P5, P6, P7, P10).

Observed behavior often differed from these claims. Although participants described LLMs as a last resort for debugging, screen recordings showed most used LLMs immediately after compilation errors. No participants used debuggers despite 43\% reporting debuggers use as part of their problem-solving strategy in the pre-survey.

\section{Discussion}
Our analysis revealed that the participants using LLMs for programming assignments often adopted what we term \textit{pseudo-apprenticeship}. The participants positioned themselves as apprentices to AI systems, relying heavily on observing expert-like solutions but skipping later stages of learning that promote progress toward independent inquiry. We present this model of learning and offer interventions for its occurrence.

\subsection{Pseudo-Apprenticeship as a Learning Pattern}

We define \textit{pseudo-apprenticeship} as a pattern in which students treat LLMs as expert models but do not progress through the stages of cognitive apprenticeship that build autonomy. Cognitive apprenticeship theory~\cite{collins1989cognitive} describes how novices develop expertise through guided practice with experts, gradually moving from observing expert performance (modeling) to receiving guidance (coaching), working with diminishing support (scaffolding), articulating reasoning, reflecting on alternative approaches, and ultimately exploring independently. Collins et al. note that the later phases of apprenticeship are essential for students to "gain conscious access to and control of their own problem-solving strategies~\cite{collins1991cognitive}," yet research on implementing these stages in computing education remains under-explored compared to the earlier methods~\cite{shah2024review}.

In our data, the students often remained in the initial modeling phase, using LLM-generated solutions as authoritative references but seldom engaging in later apprenticeship facets like scaffolded practice or articulation. This pattern reflects not only how LLMs deliver complete solutions but also how students interact with them. We observed three recurring behaviors that characterize pseudo-apprenticeship: (1) reading without refining, (2) bypassing productive struggle, and (3) ceding problem ownership, each described below.

\subsubsection{Reading Without Refining}

Participants devoted considerable time to examining AI-generated code, aiming to understand how the provided solutions worked. They often treated the generated code as authoritative, rarely issuing follow-up prompts or significantly adapting the code to their needs. This pattern aligns with the modeling phase of cognitive apprenticeship, where learners observe expert solutions. However, cognitive apprenticeship also emphasizes moving beyond observation through scaffolded practice, articulating design choices, and exploring alternative approaches. Even with varying experience levels and problem difficulty, participants seldom engaged in such refinement, leaving their learning focused on comprehension rather than transformation or independent problem-solving.

\subsubsection{Bypassing Productive Struggle}

The participants showed low tolerance for confusion or uncertainty, often seeking AI assistance at early signs of difficulty, especially in open-ended tasks or tasks with unfamiliar syntax. This avoidance bypasses the ``desirable difficulties'' known to strengthen retention and transfer by slowing apparent progress but engaging the retrieval and problem-solving processes that build robust learning~\cite{bjork1994memory}. In instructional models such as cognitive apprenticeship, supports are intentionally faded so that learners experience this productive struggle~\cite{wood1976role,hiebert2007effects,collins1989cognitive}. Immediate and complete LLM responses removed opportunities for students to engage in that struggle, particularly by preempting the need to plan their approach before coding. Similar patterns are observed in professional knowledge work, where generative AI use often reduces the cognitive effort devoted to critical engagement with tasks~\cite{lee2025impact}.

\subsubsection{Ceding Problem Ownership}

The participants often let LLM responses dictate their problem-solving approach. LLMs frequently produced complete solutions, and students rarely redirected or adapted these outputs. Even focused queries, such as brainstorming ideas or exploring specific programming constructs, often returned full solutions. This shifted interactions from adaptive help-seeking, where learners request only enough support to keep working independently~\cite{aleven2003help}, to non-adaptive help-seeking, where key decisions are ceded to the source of help. This reinforced a modeling-only pattern of cognitive apprenticeship, leaving little opportunity for productive struggle and independent inquiry.

\subsubsection{Influence of Task Characteristics}

Our findings suggest that task characteristics such as open-endedness and familiarity are factors determining whether students fall into pseudo-apprenticeship patterns. When tasks were narrowly scoped and relied on familiar concepts, as in \activity{1}, participants more often attempted independent work before consulting LLMs. These well-defined problems activated the students' existing knowledge and confidence, delaying their entry into the modeling-without-progression pattern.

\subsection{Implications for Educational Practice}

We outline strategies recommended for instructors to mitigate the pitfalls of pseudo-apprenticeship. These strategies are derived from our findings but their effectiveness requires future empirical validation.

\subsubsection{Designing for Intentional Workflows}
Assignments can be structured so that students pause and make explicit choices rather than outsourcing the solution in one step. Requiring reasoning in plain text before coding or asking students to identify gaps in their knowledge before consulting external resources can create these moments of reflection and decision-making. Making authorship patterns visible, for example by distinguishing between LLM-generated or pasted code and student-authored code in the IDE, can reinforce this reflection by giving students a clearer view of their own contributions. Students in our study expressed both apprehension about their AI use and limited insight into how much of their work originated from AI, suggesting that surfacing these patterns can build awareness and agency.

\subsubsection{Supporting Early Struggle}
Many students turned to LLMs at the first signs of difficulty, bypassing opportunities for productive struggle. Releasing assignments during class or lab, where guidance is available from peers or instructors, aligns with students' stated preference for in-person help while addressing their interest in the immediacy of AI assistance. Improving course materials so they are searchable, contextual, and explicitly linked to assignments could also reduce the appeal of one-shot AI solutions. Students valued these traditional materials in interviews but found them slower to access and less contextually relevant during the activities.

\subsubsection{Teaching the Traps of AI Use}
Because students will use LLMs regardless of policy \cite{lau2023ban}, instruction should explicitly address common traps in AI use. Students need awareness of how prompt wording can lead to overly complete solutions, how providing too much problem context can shift control away from them, and how easily their original intent can be lost. Without explicit attention to these risks, students may form habits of overreliance that prevent them from progressing beyond basic modeling behaviors.

\section{Limitations}
This study has several limitations that affect how its findings should be interpreted. First, our sample was small (14 participants) and drawn from a single institution, which limits the generalizability of our results. While our participants represented a range of experience levels within the early undergraduate curriculum, they may not reflect patterns of LLM use at other institutions, in different instructional contexts, or among students with more advanced programming experience.

Second, the programming activities were designed to approximate homework-style tasks but were not actual course assignments. This choice allowed us to control for task scope and difficulty while observing students in a single session. However, it also means that participants may not have experienced the same stakes, stress, or motivation as when completing real coursework. Time limits were imposed to simulate deadline pressure, but students may approach LLM use differently when working on authentic assignments with flexible deadlines.

Third, our observations focused on short-term interactions during a single session. We did not measure longer-term learning outcomes or track whether pseudo-apprenticeship behaviors persist or change over time. Students’ LLM use patterns may evolve as they gain experience with programming or as institutional policies and tools change.

Finally, although we used think-aloud protocols, surveys, and interviews to triangulate data, these self-reported measures can diverge from actual behavior.


\section{Conclusion}
This study reveals how CS1 students interact with large language models when permitted to use any resources during programming tasks. Through detailed observation of 14 students thinking aloud while coding, we identified a prevalent learning pattern we term \textit{pseudo}-\-\textit{apprenticeship}, where students position themselves as apprentices to AI systems but fail to progress into the developmental stages that build programming autonomy.

Our findings show that students frequently bypass planning and exploration phases, instead working backward from complete AI-generated solutions. This pattern was most pronounced when students faced unfamiliar concepts or open-ended tasks, with participants spending considerable portions of their time interacting with LLMs. While students devoted much effort to understanding AI-generated code, they rarely engaged in the refinement, articulation, or independent exploration that characterizes effective learning through cognitive apprenticeship.

The pseudo-apprenticeship pattern emerges from several factors: students' low tolerance for productive struggle, the tendency of LLMs to provide complete solutions even when not explicitly requested, and disconnects between students' stated learning goals and their actual behaviors under time pressure. These findings suggest that the availability of AI tools alters how novices approach problem-solving in ways that may limit their development as independent programmers.

For computing educators, these results highlight the need for pedagogical adaptations that preserve opportunities for productive struggle while acknowledging that students will use LLMs regardless of policy. We recommend designing assignments that make AI assistance patterns visible, creating structured moments for reflection before tool use, and explicitly teaching students about the pitfalls of over-reliance on AI-generated solutions.

As generative AI becomes increasingly integrated into programming practice, understanding how novices interact with these tools is crucial for maintaining the educational value of introductory programming courses. This work contributes empirical evidence of actual student behaviors and introduces the pseudo-apprenticeship framework as a lens for understanding and addressing the challenges of learning to program in settings where students use LLMs.


\bibliographystyle{ACM-Reference-Format}
\bibliography{bibliography}

\end{document}